\documentclass[11pt,a4paper]{article}
\pdfoutput=1

\usepackage[T1]{fontenc}
\usepackage[utf8]{inputenc}
\RequirePackage[colorlinks,citecolor=blue,urlcolor=blue]{hyperref}
\usepackage{amssymb,amsmath,amsthm,color,outlines,subfigure,comment}
\usepackage[small]{caption}
\usepackage{graphics}
\usepackage{graphicx}
\usepackage{enumitem}
\usepackage{stackrel}
\usepackage{tikz}
\usepackage{authblk}
\usepackage[square,numbers,sort&compress]{natbib}
\usepackage[margin=1.0in]{geometry}
\usepackage{fancyhdr}

\fancypagestyle{plain}{%
  \fancyhf{}%
  \fancyfoot[C]{Copyright © 2022 for this paper by its authors. Use permitted under Creative Commons License Attribution 4.0 International (CC BY 4.0).\\\textit{Proceedings of the Conference on Applied Machine Learning for Information Security}, 2022}%
}

\newcommand{\BP}{{\mathbb{P}}}
\newcommand{\BR}{{\mathbb{R}}}

\newcommand{\cD}{\mathcal{D}}

\def\baq#1\eaq{\begin{align}#1\end{align}}
\def\ban#1\ean{\begin{align*}#1\end{align*}}

\def\bredbf#1\eredbf{{\color{red}{\bf ???? #1 ????}}}
\def\BeginRedComment#1\EndRedComment{({\color{red}{\bf Comment:} {\it #1}}) }
\def\BeginBlueComment#1\EndBlueComment{({\color{blue}{\bf Comment:} {\it #1}}) }

\makeatletter
\newcommand{\opnorm}{\@ifstar\@opnorms\@opnorm}

\newcommand{\@opnorm}[2][]{%
  \mathopen{#1|\mkern-1.5mu#1|\mkern-1.5mu#1|}
  #2
  \mathclose{#1|\mkern-1.5mu#1|\mkern-1.5mu#1|}
}
\makeatother

\theoremstyle{definition} 

\title{Network Security Modelling with Distributional Data}
\author[1,2]{Subhabrata Majumdar\thanks{Corresponding author, subho@biasbounty.ai}}
\author[3]{Ganesh Subramaniam}
\affil[1]{AI Vulnerability Database}
\affil[2]{Bias Buccaneers}
\affil[3]{AT\&T Data Science and AI Research}
\date{} 

\begin{document}

\maketitle

\begin{abstract}
We investigate the detection of botnet command and control (C2) hosts in massive IP traffic using machine learning methods. To this end, we use NetFlow data---the industry standard for monitoring of IP traffic---and ML models using two sets of features: conventional NetFlow variables and distributional features based on NetFlow variables. In addition to using static summaries of NetFlow features, we use quantiles of their IP-level distributions as input features in predictive models to predict whether an IP belongs to known botnet families. These models are used to develop intrusion detection systems to predict traffic traces identified with malicious attacks. The results are validated by matching predictions to existing denylists of published malicious IP addresses and deep packet inspection. The usage of our proposed novel distributional features, combined with techniques that enable modelling complex input feature spaces result in highly accurate predictions by our trained models.
\end{abstract}


\section{Introduction}
\label{sec:intro}
Security monitoring of Internet Protocol (IP) traﬃc is an important problem that is growing in prominence. An exploding volume of internet traﬃc and a wide variety of devices connecting to the internet in recent years have contributed to the increase in malicious activity that can harm both individual devices and carrier networks. Given the lasting damage of internet security breaches \cite{chengetal17}, it is important to monitor this IP traﬃc for malicious activity and ﬂag anomalous external IP addresses that may be causing or directing this activity through communications with internal devices on a real time basis.

There are a large number of challenging statistical problems in network security. For example, there is ongoing research on the identification of various malicious events like scanning, password guessing, Distributed Denial of Service (DDoS) attacks, malware injection, and different spams attacks \cite{handaetal19}. The focus of this paper is on the detection of botnet attacks---specifically, identifying host IP addresses (also known as C2 or "Command and Control") that send instructions to infected bots (devices) on the nature of the attack to be perpetrated.

Reviewing the literature in network security, we observed that the current trend of NetFlow analytics and ML modelling is device-centric, i.e., the analysis of the internet traffic routed through a device to determine whether it contains malicious activity. For example, \citet{Evangelou2016PredictabilityON} used regression trees to model individual device behavior based on input features constructed from historic NetFlow data. In contrast, we perform a host-centric analysis, looking for host IPs (that devices are connecting to) acting in a possibly malicious way, particularly as the command and control server (C2) for a botnet. While a bot device may only have a small proportion of its traffic as malicious, the Host/C2 will have most of its traffic involved in the malicious activity, and therefore generate a stronger signature. Our analysis is aimed at looking for such signatures. 

Scanning the literature for methodology, recent papers have used supervised and unsupervised ML techniques for botnet detection. For example, \citet{tegeler2012botfinder} used ﬂow-based methods to detect botnets, \citet{choi2009botgad} detected botnet traﬃc by capturing group activities in network traﬃc, and \citet{karasaridis2007wide} developed a K-means based method that employs scalable non-intrusive algorithms to analyze vast amounts of summary traﬃc data. 
However, a number of structural challenges increase the complexity of NetFlow data, calling for sophisticated ML techniques---based on statistical insights---for the analysis and modeling of such datasets. As examples, getting high-quality training data is a known problem. Some attempts---such as the CTU project \cite{ctu13}---have been made to obtain sample data for a number of known malwares. Being able to establish ground truth is also difficult. The process of conﬁrming an IP address as a bad actor is expensive in terms of time and eﬀort, and may even require manual review by a security analyst or deep packet inspection (DPI). Finally, it is challenging to capture the distributional nature of features in raw flow traffic data. For a device or IP-level analysis, the raw data may contain multiple records, giving a {\it distribution} of values for features such as number of packets or bytes transferred that needs to be succinctly summarized before applying downstream ML techniques. In past work, \citet{Gu07,Gu08,GuNDSS} used unsupervised methods on flow and some distributional features to detect botnets in NetFlow data, and developed a scalable framework to apply such methods as a filtering step before DPI \cite{Zhang11}.

In this paper, we present a statistical pipeline to model the IP network traffic for a given day using NetFlow data and to detect botnet attacks, while minimizing the need for expensive techniques such as DPI. We summarize the flow traffic feature distributions into carefully crafted IP-level feature vectors, then feed these into supervised ML models to predict whether an IP is malicious or benign. While \citet{Gu08} has used a similar featurization method in an unsupervised setting, we take a more principled approach, guided by empirical evidence of differences in flow feature distributions of traffic through C2 vs. benign hosts \cite{https://doi.org/10.48550/arxiv.2108.08924}. To this end, we experiment with a number of ensembling strategies to combine predictions from multiple models. The best performing models in our approach are able to accurately flag malicious IPs ahead of time.


\section{Preliminaries}
We first introduce a few basic concepts, and give a high-level overview of our ML pipeline.

\subsection{NetFlow Data}

NetFlow\footnote{\url{https://www.kentik.com/kentipedia/netflow-overview}} is a network protocol developed by Cisco for collecting, analyzing and monitoring of packet capture data. A fundamental tool for characterizing IP traﬃc, NetFlow data is comprised of source and destination IP addresses, packets and bytes transferred, and duration and IP protocol number used. While there are other components of IP traffic data---such as data from HTTP log ﬁles and DNS requests---NetFlow data is easily available, allowing analysts to create a data-driven funnel of highly probable IPs for further, more intensive investigation. To this end, two reasons make it necessary to extract and craft relevant features from NetFlow data. Firstly, NetFlow data is massive in scale: for a single day, the size of flow traffic data passing through a communication network may run into several hundreds of terabytes. Secondly, NetFlow data has a limited number of attributes, as shown in Table~\ref{tab:netflowtab}. These reasons predicate the need to extract and aggregate relevant statistical features for feeding into downstream investigations. 


\begin{minipage}[b]{.49\textwidth}
\begin{center}
\begin{tabular}{ |l|l| }
  \hline
  \multicolumn{2}{|c|}{NetFlow Data Fields} \\
  \hline
  1. & Source IP address \\
  2. & Destination IP address\\
  3. & Source port \\
  4. & Destination port\\
  5. & Bytes transferred    \\
  6. & Packets transferred    \\
  7. & Start Time \\
  8. & End Time \\
  9. & IP Protocol number\\
  10. & Flag \\
   \hline
\end{tabular}
\label{tab:netflowtab}
\captionof{table}{NetFlow Data}
  \end{center}
\end{minipage}
\begin{minipage}[b]{.4\textwidth}
\begin{center}
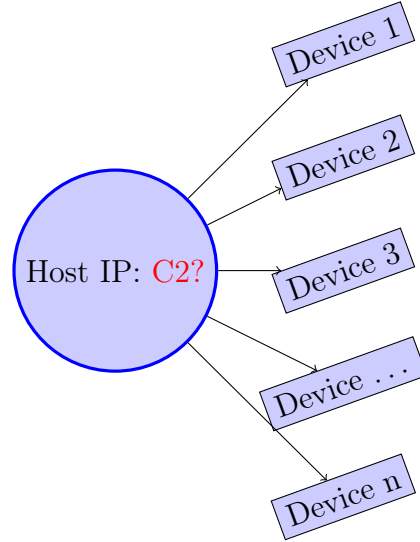

\usetikzlibrary{shapes.geometric}
\begin{tikzpicture}[fill=blue!20]
\path      (3,0) node(g) [circle,draw=blue,very thick,fill] {\large Host IP: \textcolor{red}{C2?}}
      (6,3) node(h) [rectangle,rotate=20,draw,fill] {\large Device 1}
      (6,1.5) node(i) [rectangle,rotate=20,draw,fill] {\large Device 2}
      (6,0) node(j) [rectangle,rotate=20,draw,fill] {\large Device 3}
      (6,-1.5) node(k) [rectangle,rotate=20,draw,fill] {\large Device $\dots$}
      (6,-3) node(l) [rectangle,rotate=20,draw,fill] {\large Device n};


\draw[->] (g) -- (h);   
\draw[->] (g) -- (i);
\draw[->] (g) -- (j);
\draw[->] (g) -- (k);
\draw[->] (g) -- (l);

\end{tikzpicture}
\end{center}
\label{fig:DevVsHostFlow}
\captionof{figure}{C2-Centric Traffic Flow}
\end{minipage}

\subsection{What is a Botnet?}
 
In recent years, botnets have emerged as one of the biggest threats to network security among all types of malware families, since they have the ability to constantly change their attack mechanism in scale and complexity \cite{silva2013botnets}. A botnet is a network of compromised devices called bots and one or more Command \& Control (C\&C or C2) servers. Generally speaking, the bots could be a PC, a server, an Internet of Things (IoT) device or any machine with access to the internet. In this type of threat, the orchestrator---called the {\it botmaster}---authors a malware that operates on each bot. Devices are infected with the malware in several ways, such as ``drive by downloads'' which refers to the (unintentional) download of malware as a result of visiting a website or opening an infected email. The botnet control system, i.e. the C2 server, is the mechanism used by the botmaster to send commands and code updates to bots which then conduct the attacks. Due to the prevalence of firewalls, the botmaster cannot contact devices directly. Typically, the bot malware has instructions to contact the C2 to establish the communications and to receive instructions on any attacks to be perpetrated. The nature of such attacks vary in scale and sophistication. Examples of attacks by botnets include transmitting malware, using the bots to perform diﬀerent illegal activities, e.g. spamming, phishing, or stealing conﬁdential information, and orchestrating various network attacks (e.g. DDoS) \cite{fi13080198}. 

\subsection{ML pipeline for botnet detection}
As noted in Section~\ref{sec:intro}, the most common approach to identifying botnets is to look at individual devices and analyze their traffic with various hosts. This means analyzing each of the device’s connections, as shown in the right panel of Figure~\ref{fig:DevVsHostFlow}, for possible malicious traffic. However, the connection traffic between a bot device and the C2 may not look significantly different than other benign traffic for that device, and/or or be a small portion of its traffic. Therefore, each connection must be analyzed individually. 

In this paper, we take a C2-centric view of the data instead, analyzing the external host for C2 behavior. Figure~\ref{fig:DevVsHostFlow} shows the traffic between one external host IP address and several devices internal to a carrier network, each having a distinct IP address. We aggregate device traffic for each such (external) host IP address, and use this aggregate data to answer the question: {\it which host IP addresses have traffic that looks like a \textit{botnet command and control (C2)} pattern?} Assuming that a C2 server aims to control a large number of bot devices, such controlling behavior will manifest in the interaction data between a C2 server and its paired devices. Therefore, we can look for the C2 signature as the predominant traffic pattern over all the paired devices. This C2-centric approach allows for richer aggregation, and fewer samples that need to be analyzed, thereby improving the accuracy and scalability of the resulting detection. We construct features for each host IP from the NetFlow data traffic between the host and all of its associated device IPs (see right panel in Figure~\ref{fig:DevVsHostFlow}), then train a machine learning (ML) models using the constructed features to predict a Host IP address as malicious or benign.
\begin{figure}[t]
    \centering
    \includegraphics[width=\textwidth]{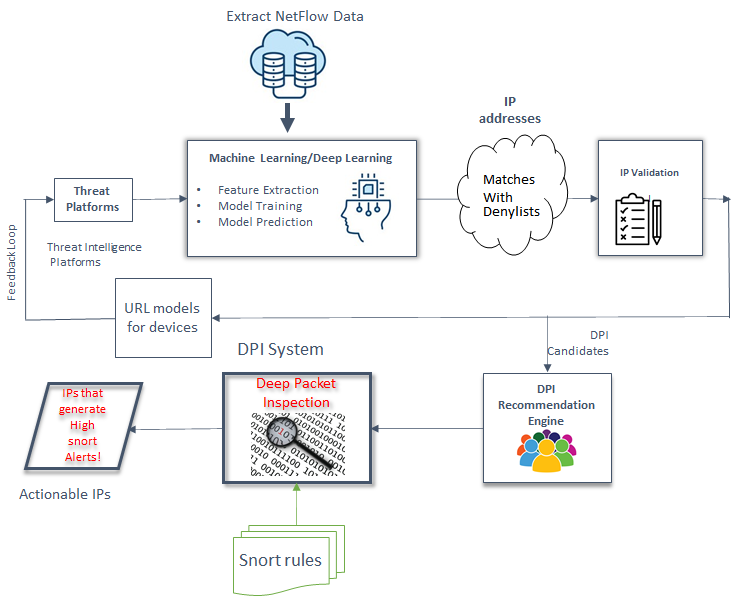}
   \caption{The ML Botnet Pipeline}
    \label{fig:framework}
\end{figure}

Figure~\ref{fig:framework} shows a simplified view of our pipeline. The threat platform monitoring activity in an internet network ingests NetFlow data from multiple traffic domains on a daily basis. We use ML models trained on historical data to predict the label for external host IP addresses. These models detect IP addresses associated with Botnet C2, Trojans, and other dangerous malware families. As discussed in Section~\ref{sec:intro}, a major challenge in network security is knowledge of ground truth, i.e. proof that IP addresses predicted by the ML model as malicious are indeed malicious Botents. A na\"ive approach would be to see if these predicted IPs show up in existing lists of malicious IPs in threat platforms, i.e. denylists. The problem with this approach is that the sources of denylists are diverse, ranging from crowdsourced information to analysis of malware samples. Such sources are of variable and not always reliable quality. Due to this limitation, we take a multi-tiered approach in validating the model-generated list of potentially malicious IPs. We first use the network denylists to filter out any known malicious IPs, then pass the remaining list of IPs to a validation engine. The validation engine mechanizes rule-based review processes typically done manually by security analysts. The shorter list passed on from the validation engine finally goes through deep packet inspection (DPI)---the most definitive validation process---to see if these IPs generate alerts associated with well known vulnerabilities (IPs subject to legal review). The final list of IPs coming out of this process tagged as malicious are considered as `actionable' IPs, ready to be utilized in future security use cases. The above three-step process not only saves resources by successively filtering an initial list of suspicious IPs before using the expensive DPI validation technique, but also eliminates possible `false positives'---IPs that may either be allowlisted or sinkholes, or belong to well known content delivery networks and cloud providers.

\section{Materials and Methods}
In this section, we give details of the NetFlow data analysis and model building. We start with feature engineering, then describe the modelling steps.

\subsection{Data}
We used daily flow data processed from numerous traffic domains associated with several classes of service for a telecommunications solution provider. Based on the matches from the native threat intelligence platform that maintains a list of confirmed IP addresses belonging to several malicious botnet families, some IP addresses are labeled `malicious', and others as `unknown'. To construct the above list, active malware sample traffic traces observed in the network within past 30 days were used. As the list of IP addresses associated with the malicious families is very small compared to the entire traffic, there was a significant imbalance of class labels. To mitigate the class imbalance, we sampled 1000 IP addresses from the `unknown' class for every day of the month of December 2021, took all IP addresses associated with the `malicious' class, and used the traffic flowing through these IPs to construct our training dataset. This hand constructed training data had $\sim $17\% malicious and 83\% unknown traffic. All traffic from the subsequent month (January 2022) were used as the test dataset.

\subsection{Feature engineering}

The first important step for building a predictive model for detecting botnets is the exploration of the input feature space. We need to craft input features that  enable sufficient description of the NetFlow traffic to distinguish between malicious and benign IPs.  The ability of the engineered feature space to provide pertinent information is critical to the subsequent ML step, as the underlying assumption of the classification models used is that feature characterization of the malicious botnet and benign NetFlow traffic have different distributions. In previous exploratory work on NetFlow data \cite{https://doi.org/10.48550/arxiv.2108.08924}, we discovered traffic traces associated with known botnet families, in other words `live' botnet traffic, i.e. C2 IP addresses.  Such IP addresses are called malware samples. Using the flow data of IP addresses from these malware samples we uncovered a number of main characteristics or signatures that differentiate normal traffic from botnet traffic. \citet{https://doi.org/10.48550/arxiv.2108.08924} presented a comprehensive discussion of feature engineering for NetFlow data to help build an informative feature set for botnet prediction using ML.

This set of features can be categorized into two majot groups: (1) flow size features, (2) beaconing features. In addition to them, in this paper we also  use distributional features for flow variables to encode granular information on flow variable distributions.

\subsubsection{Flow size features}
The first set of statistical features engineered from cthe NetFlow data are based on flow sizes, that indicate the total number of bytes/packets transferred between the source and the destination endpoints for a given flow. In our case, this is the traffic between a single source IP (SIP) and all the devices it communicates with, i.e., aggregate traffic between a SIP and all devices. Our exploratory analysis of flow sizes using live botnet data indicate the flow size characteristics for C2 servers are significantly different from flow size characteristics for benign servers. The differences can be attributed to several factors, the main one being the botnet traffic tries to
maintain a low profile to avoid detection. As a result, botnet flow feature values are usually small, and have minimum variation across time. In contrast, benign flows show more diversity in flow sizes, i.e., assume a wide range of values. Other statistical features we use comprise of bytes, packets, duration, bytes to packet ratio, byte and packet rates. Finally, it is possible to infer who initiated the connection--the external host IP or the device---using port information. Thus we include one-hot encoded port indicators as input features.

\subsubsection{Beaconing features} 
Malware downloaded by compromised internal devices or servers displays {\it beaconing} behavior, which involves sending short and routine communications to the C2 server. Beaconing signals that the infected device in the internal network is now available and listening to the C2 server for further instructions. We developed a number of features to specifically detect the presence of beaconing activity to confirm that the signaling is active. As an example, from the observed  sequence of source IP start times, the  \textit{inter-arrival times} are defined as the differences between start times of successive flows. If inter-arrival times display a periodic pattern, then beaconing signal is present. If they are random, then beaconing signal is not present. Based on such logics, several statistics computed based on the set of inter-arrival times form the basis of beaconing features.

\subsubsection{Distributional features}

As indicated earlier, the rationale for our IP-level analysis is the hypothesis that C2 servers demonstrate markedly different flow behavior compared to benign IP addresses. Translating to statistical terms, this means that the distributions of flow features for the two classes are very different. Only using static summary statistics of these distributions such as mean, median, or standard deviation (as used in the engineered flow and beaconing features above) may not be sufficient to optimally tell apart malicious and benign IPs. Because of this reason, we craft an additional number of features, from {\it quantiles} of IP-level raw flow feature distributions. 

As an example, consider the three input features such as packets, bytes, packets-to-bytes ratios have multiple observations per IP. Denote their distributions for a device as $\cD_p, \cD_b, \cD_r \in \BP$, respectively. Here $\BP$ is the set of all real-valued probability distributions. Assume that whether an IP is malicious or not is a function of these distributions:
$$ \mathbb I( \text{malicious}) = f(\cD_p, \cD_b, \cD_r). $$

This model may be approximated using summary statistics such as mean $\mu(\cdot)$ and standard deviation $\sigma(\cdot)$ of an IP-level feature distribution:
$$ \mathbb I( \text{malicious}) \simeq f((\mu(\cD_p),\sigma(\cD_p)),
(\mu(\cD_b),\sigma(\cD_b)), (\mu(\cD_r),\sigma(\cD_r)). $$
In addition to the above somewhat simplistic feature summaries, we use a wider spectrum of distributional features, obtained using quantiles of each feature distribution:
$$ \mathbb I( \text{malicious}) \simeq f(G(\cD_p), G(\cD_b), G(\cD_r)), $$
where $G \equiv (\mu, \sigma, Q); Q: \BP \mapsto \BR^n$ indicating the vector transformation giving $n$ pre-defined quantiles from a distribution, While the transformation $Q(\cdot)$ can be made as arbitrarily high-dimensional by taking closely situated quantiles, we found that for our dataset, model performances plateau at less than than 5\% granularity of quantiles. Consequently, we set $n=20$, i.e. consider 5\%, 10\%, …, 95\%, 100\% quantiles of the respective flow feature distribution.

Generally, there is significant overlap in feature-level summary statistics across malicious and benign IPs. Using a larger number of quantiles that adapt to the shape of a distribution allows us to tease out the differences between these two classes more accurately. Moreover, some summary statistics such as standard deviation require a large enough sample size for the calculated value to be usable. As a result, IPs with smaller number of observations may be dropped from the analysis and/or modeled inaccurately. Quantile-based features do not have this limitation.

\begin{figure}[t]
    \centering
    \includegraphics[width=.4\textwidth]{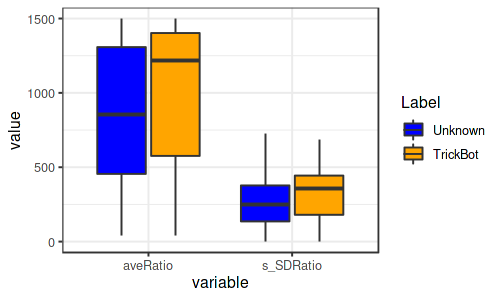}
    \includegraphics[width=.95\textwidth]{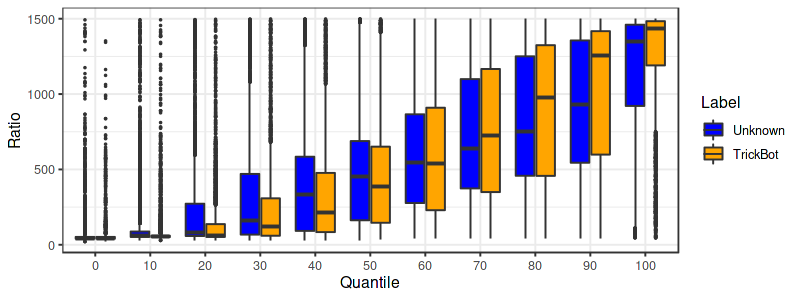}
    \caption{Comparison of static (top) vs. quantile (bottom) features for packets-to-bytes ratio.}
    \label{fig:compare}
\end{figure}

As an example, consider the plot of the average and standard deviations of packets-to-bytes ratio for IPs that are known to be malicious, vs. IPs with unknown status, 
and compare it with the plot of IP-level deciles (quantiles at 10\% intervals) of the same feature (Figure~\ref{fig:compare}). Even though values in the two classes intersect heavily for the static summary statistics, quantile features are different for samples across the classes in lower and higher quantiles.  

\subsection{Ensemble Models for NetFlow Data}

ML techniques---both supervised and unsupervised---are widely used in cybersecurity. Typically, unsupervised techniques are used on known malware samples (e.g., live botnet traffic) to explore statistical features of the malware. Supervised ML models are ideally trained on high-quality data containing reliable labels (malicious vs benign). The general principle is that as the training data is labeled, the model will `learn' from the labeled patterns to build the classifier which the can be used it to predict class labels for IPs in new traffic data.

Many researchers have employed such models for botnet detection. Here the traffic of IP addresses associated with known set of botnet families are used as training data to learn a prediction function that classify an IP address as botnet C2 server or benign. For example, \citet{https://doi.org/10.48550/arxiv.2108.08924} built predictive models using Random Forest and deep learning techniques on flow size and beaconing features and demonstrated the predictive performance for these models for a one month period. 

In this paper, we investigate two novel ideas:
\begin{enumerate}[nolistsep,leftmargin=*]
    \item The use of distributional features in the models in addition to the traditional flow features,
    \item The use of ensemble methods \citep{Opitz1999,gneiting2005weather} by combining the predictions of multiple (weak) prediction methods (bucket of models).
\end{enumerate}

Ensemble models combine or stack several base models together. They are aimed at maximizing the contribution from diverse models to get a wider understanding of the class-distinguishing input features, especially for complex datasets. In theory, ensemble models can improve both the accuracy and stability of predictions over individual models by taking advantage of the underlying differences and strengths of the base models. Ensemble methods are used extensively in other fields like medicine (e.g. MRI datasets \cite{mri}), finance (fraud detection \cite{fraud}), image analysis (Face recognition \cite{imgrecog}) and meteorology \cite{gneiting2005weather}, to name a few.

      

We use a number of ensembling starategies. Firstly, we use ML models that are ensembles by definition: random forest \citep[RF]{breiman2001random} on the base features, RF on PCA-transformed input features, and two versions of boosting methods \citep{friedman2001greedy}: gradient boosting and XGBoost. Secondly, we use two simpler approaches:  Logistic regression (linear classifier), LASSO (regularization), and stack all six of our models using a GLM-based ensembling strategy \cite{caretens}. Thus, we cover different types of ensembling strategies: parallel combination (RF models), serial combination (boosting models), and a stacked combination of all models.

\section{Evaluation}
We did all analyses using the \texttt{R} statistical software, utilizing the packages \texttt{caret} \citep{JSSv028i05} and \texttt{caretEnsemble} \cite{caretens}, and a 10 fold cross-validation for hyperparameter tuning. For model performance evaluation, we obtain boxplots of performance metrics using bootstrapped samples from the test data (resample size 1000).

\begin{figure}[t!]
    \centering
    \includegraphics[width=.9\columnwidth]{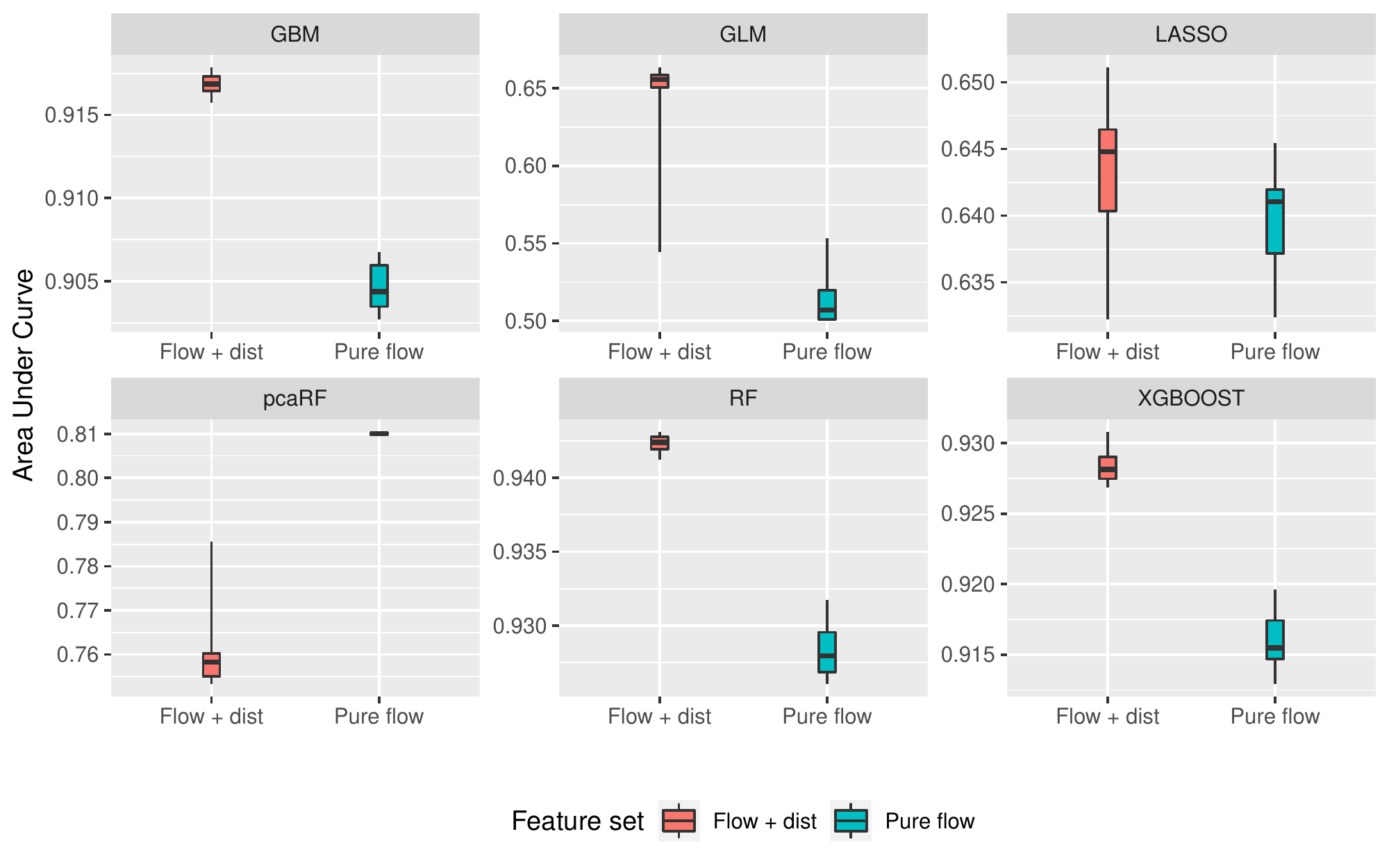}
    \includegraphics[width=.9\columnwidth]{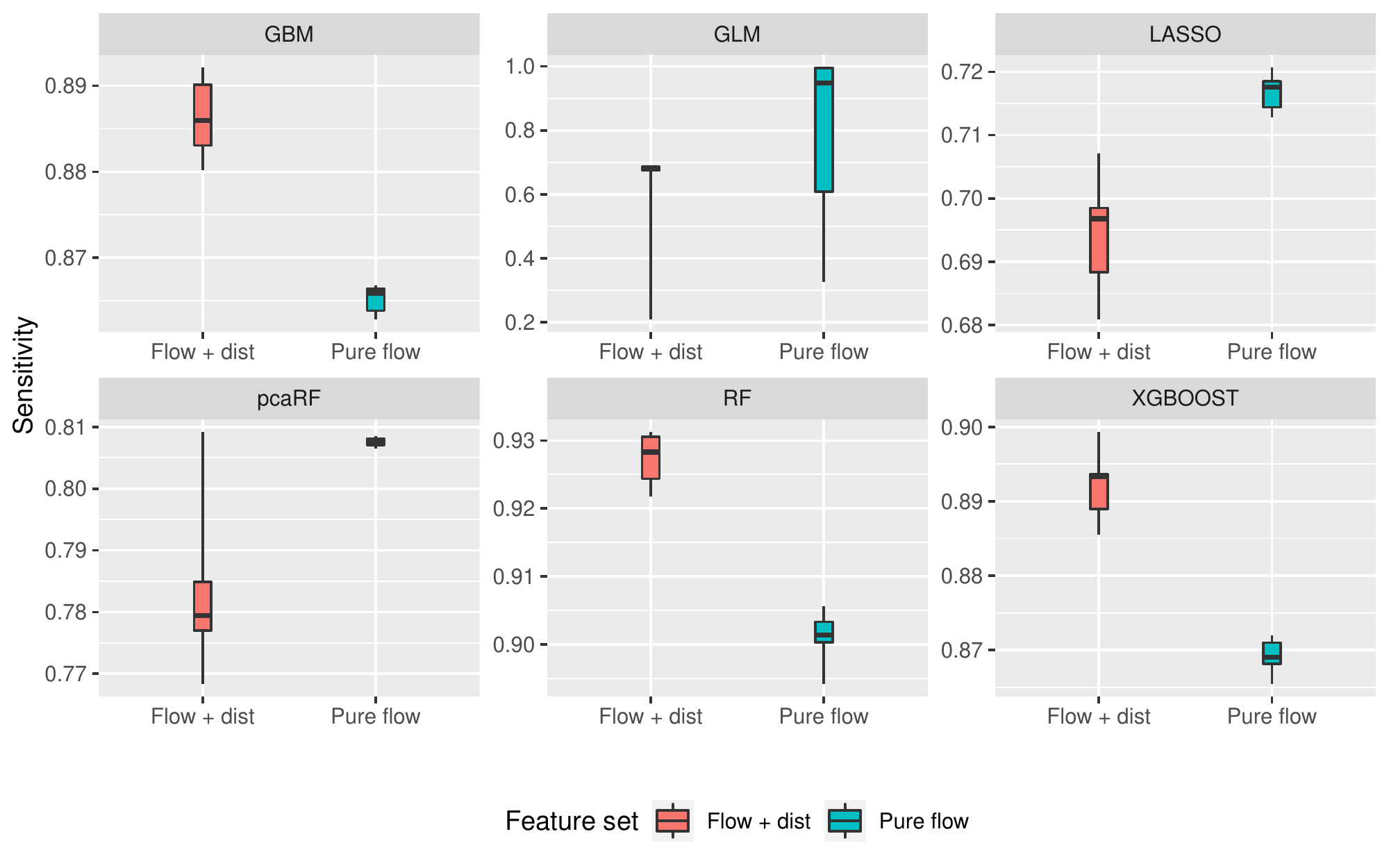}
    \caption{Comparison of model performance using AUC (top) and sensitivity (bottom).}
    \label{fig:ensemble_compare}
\end{figure}

Figure~\ref{fig:ensemble_compare} presents two performance metrics---Area Under the ROC curve (AUC) and sensitivity---for each of our six base models. We compare two feature sets for model evaluation: the conventional flow features related to flow size and beaconing, without and with our novel distributional features for bytes, packets, and bytes to packets ratio. Random forest and the boosting methods performed well---XGBoost being the fastest in terms of computation time as well---whereas the linear models performed poorly. Random forest on PCA transformed input features did better than the GLMs. We generally see the inclusion of quantile-based flow features improving model performance for both metrics---the effect being stronger for AUC. The positive effect of distributional features is pronounced across both the metrics on the three best performing models: RF, GBM, and XGBoost. Finally, stacking the predictions from the different classifiers using a simple linear model, we get an improvement in the performance metrics: the average AUC across bootstrapped test sets for the stacked ensemble of all 6 models was 0.95, and the sensitivity was 0.83. The AUC is comparable to the better performing (RF and boosting) models, while sensitivity is not as good, potentially because of the simpler models (GLM, Lasso, pcaRF) having lower values for this metric.

\begin{figure}[t!]
    \centering
    \includegraphics[width=.9\columnwidth]{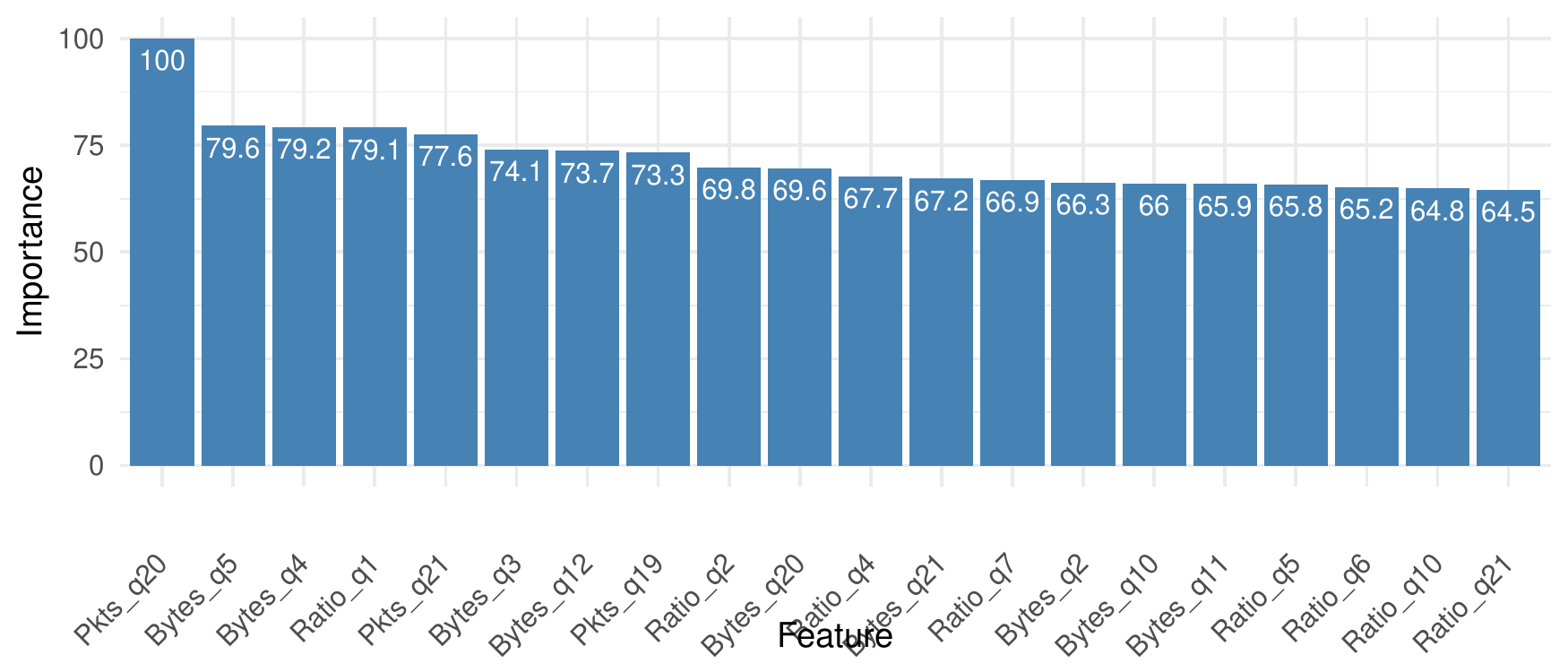}
    \caption{Top 20 important input features in the all-feature RF model. Variable importances are in percentile scale.}
    \label{fig:varimp}
\end{figure}

To understand what features are important behind behind the classification of IP addresses, we look at the variable importance plot of the all-features random forest model (Figure~\ref{fig:varimp}). The fact that all the top important features belong to the quantile feature set underline their informativeness in our IP classification scenario. We also observe that most of these top 20 quantiles are correspond to either of the tails: 14 of the 20 quantiles lie outside the Inter-quartile range, i.e. 25th and 75th quantiles.

\section{Conclusion}
Detecting botnet and other malware activity is a challenging task even with good quality labeled data sets. The framework used in this paper, namely utilizing a combination of distributional characteristics of NetFlow variables in conjunction with the stacking of multiple ML models provides a useful strategy for detecting malicious activity in IP traffic. The advantages of distributional variables are two fold: the ease of computing them versus traditional NetFlow variables, and the flexibility of choosing more or less quantiles based on computational constraints. Some NetFlow features involve computing volumes in both directions, i.e., originating and terminating directions. These are computationally expensive. The ensemble method using GLM, which is used in the context of botnet detection for the first time improved the accuracy performance and provided stability for the predictions as well. Results confirm that the Super Learner (stacked model) provides better accuracy than any of the individual models.
 
In future work, deep learning methods needs to be evaluated in comparison with `traditional' ML models for the current task. Further investigations are necessary to determine the performance of more complex ensembling methods, such as Bayesian model averaging. Our current labeled data contains a diverse mix of various malware families. It would be of interest to train separate ML models for specific families of malware samples and identify family-specific flow features instrumental behind the respective prediction models. Finally, existing research on adversarial tactics to fool statistical malware detection methods based on flow data \cite{rigaki,wright09} provide motivation to perform similar analyses of our featurzation technique and devising predictive models robust to such adversarial attacks. 

\section{Acknowledgement}
We thank Robert Archibald, Richard Hellstern, and Craig Nohl (AT\&T) for providing advice, helpful comments and technical expertise related to Network Security.

\bibliographystyle{abbrvnat}
\bibliography{arxib-main}

\end{document}